\documentclass[12pt]{article}
\usepackage{epsfig}

\usepackage{amssymb}
\usepackage{amsfonts}

\usepackage{color}
 
\def\be{\begin{equation}}
\def\ee{\end{equation}}
\def\ba{\begin{array}{c}}
\def\ea{\end{array}}

\def\ben{$$}
\def\een{$$}

\newcommand{\bea}{\begin{eqnarray}}
\newcommand{\eea}{\end{eqnarray}}

\newcommand{\bbr}{\br\!\br}
\newcommand{\kkt}{\kt\!\kt}

\newcommand{\kt}{\rangle}
\newcommand{\br}{\langle}

\begin{document}

\begin{center}

{\Large \bf

Twin Hamiltonians, \textcolor{black}{alternative parametrizations} of the  Dyson maps, and the
probabilistic interpretation problem in quasi-Hermitian quantum
mechanics

}

%
%

\end{center}

\vspace{0.4cm}

\begin{center}

  {\bf
Aritra Ghosh}$^{a,}$\footnote{present address:
School of Physics and Astronomy, Rochester Institute of Technology,
Rochester, New York 14623, USA},
  {\bf
Adam Miranowicz}$^{b}$, and
  {\bf Miloslav Znojil}$^{c,d,}$\footnote{corresponding author,
{e-mail: znojil@ujf.cas.cz}}

\end{center}


$^{a}$
School of Basic Sciences, Indian Institute of Technology, Bhubaneswar,
Argul, Jatni, Khurda, Odisha 752050, India

$^{b}$ Institute of Spintronics and Quantum Information, Faculty
of Physics, Adam Mickiewicz University, Pozna\'{n} 61-614, Poland

 $^{c}$
{The Czech Academy of Sciences,
 Nuclear Physics Institute,
 Hlavn\'{\i} 130,
250 68 \v{R}e\v{z}, Czech Republic}

 $^{d}$  {Department of Physics, Faculty of
Science, University of Hradec Kr\'{a}lov\'{e}, Rokitansk\'{e}ho 62,
50003 Hradec Kr\'{a}lov\'{e},
 Czech Republic}



\subsection*{Abstract}


In quasi-Hermitian quantum mechanics (QHQM) of unitary systems, an
optimal, calculation-friendly form of Hamiltonian is generally
non-Hermitian, $H \neq H^\dagger$. This makes its physical
interpretation ambiguous. Without altering $H$, this ambiguity
{can be}
resolved
{either via a transformation of} $H$
into {its isospectral} Hermitian form
via {a so-called} Dyson map $\Omega: H
\to \mathfrak{h}$,
{or via {a} (formally, equivalent) specification of}
a nontrivial {physical}
inner-product metric $\Theta$ in
Hilbert space.
Here, we focus on the {former} strategy.
Our {present}
construction of the Hermitian isospectral
twins $\mathfrak{h}$ of $H$
{is exhaustive. As a byproduct, it}
does not only restore the conventional
correspondence principle between quantum and classical physics, but
it also provides a framework for {a systematic}
classification of
all
{of the}
admissible probabilistic interpretations of quantum systems
{using a preselected $H$}
in
QHQM framework.


\subsection*{Keywords}.

non-Hermitian quantum mechanics;

isospectral
pairs of a non-Hermitian and
Hermitian Hamiltonian;

ambiguity of the
unitary-evolution scenarios;

classification scheme;

\newpage

\section{Introduction}

In the majority of textbooks on quantum mechanics (see, e.g., Ref.
\cite{Messiah}), the realistic Hamiltonians of unitary systems
are usually considered in their most conventional self-adjoint
form
 \be
 \mathfrak{h}= \mathfrak{h}^\dagger = \mathfrak{h}_{\rm (free\ motion)}
 + \mathfrak{h}_{\rm (interaction)}\,.
 \label{decoclass}
 \ee
In 1956, Freeman Dyson  \cite{Dyson} found such a restriction
unnecessary and, for practical purposes,
possibly even counterproductive.
Still, he did not leave the standard theoretical framework of
textbooks. Primarily, his (implicit) criticism of the
specification of quantum dynamics using Eq.~(\ref{decoclass}) was
technical. He fully
accepted the necessity of maintaining the relationship --
known as the principle of correspondence -- between quantum and
classical phenomenological models of experimentally accessible
physical reality.

For the specific many-particle system considered in {\it loc.
cit.}, the first component
 \be
 \mathfrak{h}_{\rm (free\ motion)}=
 \mathfrak{h}_{\rm (kinetic\ energy)}
 \label{simona}
 \ee
of the Hamiltonian was simply the sum of Laplacians, while
the second component was also entirely conventional, consisting of
local two-particle interaction potentials. What was truly
revolutionary, however, was the transformation
 \be
 \mathfrak{h}\ \to H = \Omega^{-1}\,
 \mathfrak{h}\,\Omega\, ,
 \label{simn}
 \ee
i.e., a tentative isospectral simplification of the preselected
{Hermitian
but user-unfriendly} Hamiltonian
{$\mathfrak{h}$}, where
 \be
 \Omega^\dagger\Omega=\Theta \neq I\,.
 \label{genn}
 \ee
This means that the author omitted the usual requirement of unitarity
of the
transformation.

The price was the manifest
non-Hermiticity of the upgraded,
{user-friendlier} Hamiltonian
{$H$},
 \be
 H \neq H^\dagger = \Theta\,H\,\Theta^{-1}\,.
 \label{Dys}
 \ee
Such a form of non-Hermiticity (called ``quasi-Hermiticity'' in
Ref. \cite{Geyer}) is merely an inessential consequence
of the non-unitarity of the invertible mapping $\Omega:
\mathfrak{h}\ \to H$ [cf. Eq.~(\ref{genn}) and also
Appendix A below]. Naturally, both of
the isospectral twin Hamiltonians $\mathfrak{h}$ and $H$ still
carried the same information about physics, i.e., about the shared
and measurable bound-state energies.

A decisive advantage of the work with twins (i.e.,
with the Hermitian but complicated $\mathfrak{h}$ and,
simultaneously, with the
non-Hermitian but user-friendlier $H$) lied in the enhanced
freedom of the description of dynamics which could vary with the
changes of the nontrivial inner-product metric $\Theta$ or
with its optional Dyson-map factor $\Omega$
(cf.  \cite{Carl,SIGMA,Ali,Dor}).
For an efficient evaluation of the experimentally
relevant low-lying spectrum of the system characterized by a
uniquely preselected self-adjoint Hamiltonian $\mathfrak{h}$, no
conceptual problems arose with the
ambiguity
of $H=H(\Omega)$, since both the reality of the
spectrum and a consistent, unambiguous probabilistic
interpretation of the system were guaranteed by
ansatz~(\ref{decoclass}).

The flexibility
of $\Omega$ have proven extremely useful in
practice -- cf., e.g., its numerous subsequent applications to
interacting-boson systems in nuclear physics \cite{Jenssen}, or,
many years later, its role in
relativistic quantum mechanics \cite{AliKG}, in
thermodynamics \cite{NIPa},
in quantum information \cite{nogo}
and even beyond the realm of quantum
physics \cite{SSH, Dysonb}.
In the original Dyson's
calculations of Ref.~\cite{Dyson}, in particular,
this flexibility
led to an
efficient inclusion of interaction-induced two-particle
correlations, markedly accelerating convergence, and markedly
enhancing the
overall efficiency of otherwise routine variational approaches in
many-body phenomenology (cf.~\cite{SSHb}).

A few years later, the problem of providing a consistent
probabilistic interpretation of non-Hermitian bound-state systems
re-emerged, following several independent discoveries showing that
certain new non-Hermitian---but computationally
convenient---Hamiltonian candidates could produce real spectra
even without any initial reference to Eq.~(\ref{simn})
(cf., e.g., Refs. \cite{Geyer, BG, DB, BB}).
{Unfortunately,
the price to pay was nontrivial:
The reality of the spectrum became fragile \cite{fragile},
leading to a}
deep conceptual distinction of the latter,
{newer model-building}
approach
{(admitting
the coexistence of {\em both\,} the bound and resonant states)}
from the Dyson's
original pragmatic strategy
{(aimed just at the
bound states with the strictly real spectra)}.

{Indeed, the main innovation
had to be seen in the possibility of
having
just a conditional form of
the reality of the spectrum
(more comments on this point will be added in
the dedicated subsection
\ref{ccvni}
below).}
{In our present paper, still,
we will restrict our attention to the
non-Hermitian models in which the reality of the
energy spectrum is ``robust'',
i.e.,
in which one only needs to find
an appropriate probabilistic interpretation of
bound states.}
Temporarily, we will denote
the
{underlying specific}
non-Hermitian Hamiltonians
{(forming just a ``strictly bound-state'' subset
of the whole ``fragile'' family)}
by the tilded symbol
$\widetilde{H}$, therefore.

We decided to use such a notation
in order to emphasize more clearly
that
the related innovative models and studies reopened
multiple methodological questions regarding the meaningful
physical interpretation of the specific
tilded-Hamiltonian models.
Their
characteristic feature
is that
the theory is
``narrowed'' and considered without any direct reference to the
isospectrality
mapping between the preselected non-Hermitian
candidate $\widetilde{H}$ for the Hamiltonian
(in Schr\"{o}dinger picture) and its
manifestly Hermitian twin
$\widetilde{\mathfrak{h}}$.
Indeed, such a loss of
correspondence (\ref{simn})
has to be interpreted as
a loss of a clear specification of the
underlying physics, i.e., as
one of
the most significant weaknesses of the
non-Hermitian extension
of the Schr\"{o}dinger picture.

In this setting it is a bit unfortunate that
one of the most fundamental and obvious simplifications
of the interpretation questions
as provided
by the transition to
Heisenberg picture
(cf., e.g.,
\cite{Heissen,BC}))
is known to lead to certain truly
serious technical complications.
In
the present paper,
therefore, we intend to
keep using the
Schr\"{o}dinger-picture
philosophy.
We will
formulate, re-analyze and resolve some of
the Schr\"{o}dinger-picture-related
interpretation issues in necessary detail.

\section{A Dyson-inspired strategy\label{vni}}

\subsection{An elementary example\label{eevni}}

A few basic formal features of the replacement of
an otherwise unsuitable ``realistic'' (Hermitian, lower-case, untilded)
Hamiltonian $\mathfrak{h}$ by its ``more computationally
friendly'' (but non-Hermitian)
isospectral twin $H$ can be most clearly illustrated using a
schematic quantum system in a finite, two-dimensional Hilbert
space, i.e., in an $N$-dimensional space ${\cal H}^{(N)}$ with
$N=2$.

For a simple illustration, we take the initial Hermitian
reference Hamiltonian to be the following real and symmetric $2\times2$
matrix
\begin{equation}
\label{eq:dimer-h}
\mathfrak{h} = \omega \sigma_x
=
\left (
\begin{array}{cc}
0 & \omega \\
\omega & 0
\end{array}
\right )
\end{equation}
with two real eigenvalues $E_\pm = \pm\omega$.

One must now select a methodically instructive form of the Dyson
map. Within the Dyson-inspired model-building framework,
it is common to choose its simplest possible form (cf., e.g.,
review \cite{Ali}). For the purposes of the present illustration,
we therefore adopt a $2\times2$ matrix ansatz
\begin{equation}
\label{eq:Omega-dimer}
\Omega = e^{\frac{\alpha}{2}\sigma_y},
\qquad \alpha\in\mathbb{R}\,.
\end{equation}
Since $\sigma_y$ is Hermitian
and satisfies $\sigma_y^2={I}$, one has
\begin{equation}
\Omega^\dagger = \Omega, \qquad
\Omega^{-1} = e^{-\frac{\alpha}{2}\sigma_y}\,,
\ \ \ \
\label{HermiD}
\Theta = e^{\frac{\alpha}{2}\sigma_y}e^{\frac{\alpha}{2}\sigma_y}
= e^{\alpha\sigma_y}.
\end{equation}
Because the eigenvalues of $\sigma_y$ are $\pm 1$, the
eigenvalues of $\Theta$ are
\[
\lambda_\pm = e^{\pm\alpha} > 0\,.
\]
This means that $\Theta$ is strictly positive definite and defines
a new physical inner product in ${\cal H}^{(2)}$,
\begin{equation}
(\psi,\phi)_{\mathrm{phys}} =
\langle\psi|\Theta|\phi\rangle.
\end{equation}
Next, applying the Baker--Campbell--Hausdorff formula, one can
readily verify that
\begin{equation}
\label{eq:dimer-conj-x}
\Omega^{-1}\sigma_x\Omega
= \sigma_x\cosh\alpha + i\sigma_z\sinh\alpha\,,
\ \ \ \ \
\label{eq:dimer-conj-z}
\Omega^{-1}\sigma_z\Omega
= \sigma_z\cosh\alpha - i\sigma_x\sinh\alpha\,.
\end{equation}
Should we identify
\begin{equation}
\kappa = \omega\cosh\alpha,
\qquad
\gamma = \omega\sinh\alpha\,,
\end{equation}
and, vice versa,
\begin{equation}
\omega^2 = \kappa^2 - \gamma^2, \quad \quad \tanh\alpha
= \frac{\gamma}{\kappa}\,,
\label{viceve}
\end{equation}
one may recall Eq.~(\ref{simn}) and find that
\begin{equation}
H
= \Omega^{-1} \mathfrak{h} \Omega
 = \omega \Omega^{-1}\sigma_x\Omega \nonumber\\
= \omega\bigl(\sigma_x\cosh\alpha + i\sigma_z\sinh\alpha\bigr)
=
    \kappa\sigma_x + i\gamma\sigma_z\,.
\end{equation}
This defines a new Hamiltonian, which can be expressed in its
final matrix form,
\begin{equation}
H =
\left (
\begin{array}{cc}
 i\gamma & \kappa \\
 \kappa  & -i\gamma
\end{array}
\right )\,,\ \
\qquad \gamma,\kappa \in \mathbb{R}\,.
\label{eq:d}
\end{equation}
One immediately recognizes that the resulting matrix is
non-Hermitian but $\mathcal{PT}$-symmetric (parity-time symmetric;
see, e.g., Refs. \cite{Carlbook, Mousse, Christodoulides} for
broader physical context): {In
the related literature, some authors
prefer a replacement of the term ``$\mathcal{PT}$-symmetric
Hamiltonian''
by the equivalent expression  ``$\mathcal{P}$-pseudo-Hermitian
Hamiltonian''
or, briefly, ``pseudo-Hermitian
Hamiltonian''
(for details, interested readers
should read a thorough account of this terminological issue, say,
in \cite{Ali}).}

{What is important for us is that at
$\gamma = \kappa$, our matrix $H$}
exhibits the
so-called exceptional-point singularity (EP, see Ref.
\cite{Kato}), at which the system loses its observability. This
phenomenon has been described as a manifestation (or as a
witness) of a ``quantum catastrophe'' \cite{Bibi} or of a ``quantum
phase transition'' \cite{Swansonb}, arising from the
characteristic \cite{Berry, Heiss, Heissb}
simultaneous degeneracies of both the eigenvalues and
the eigenvectors.

Further discussion
of this theoretical and phenomenological peculiarity can be found,
for example, in
dedicated papers \cite{St} -- {\cite{EP3b}}.
In this context, it is worth noting that the transition from the
Hermitian toy-model Hamiltonian (\ref{eq:dimer-h}) to its
non-Hermitian counterpart (\ref{eq:d})
enabled not only a
consistent theoretical interpretation of the corresponding
spectrum
but also its experimental realization.
Indeed, the non-Hermitian matrix (\ref{eq:d}) has found
a widely adopted implementation in classical optics,
where it describes, e.g., a $\mathcal{PT}$-symmetric optical
dimer \cite{Mousse}.

\subsection{A remark on broader context\label{ccvni}}

{Marginally,
let us remind the readers that our above-outlined analysis
started from the bound-state hypothesis,
i.e., from the
assumption that our initial Hermitian toy-model
matrix (\ref{eq:dimer-h}) is real.
The ${\cal PT}$-symmetry of
its isospectral avatar (\ref{eq:d})
remained unbroken \cite{Carl}.
Nevertheless, one might also decide
to start from
a manifestly non-Hermitian toy-model
of Eq.~(\ref{eq:d}). Then, we immediately
reveal that its non-Hermiticity
may be enhanced
beyond the constraint of Eq.~(\ref{viceve}).
As a consequence,
the ${\cal PT}$-symmetry becomes spontaneously
broken \cite{Carl,Ali}.
After such a generalization,
the closed quantum system of our present interest
(with a real $\omega$ in Eq.~(\ref{eq:dimer-h}))
ceases to exist. As long as
the spectrum ceases to be real,
one has to change the perspective and
speak about
another, ``open'' quantum systems in which the states
acquire the physical meaning of resonances \cite{BG}.}

{It is worth adding that in
a broader physical context
the latter dynamical regime characterized
by the spontaneously broken ${\cal PT}-$symmetry
inspired new theoretical as well as
experimental efforts and developments.
Let us only mention here
the innovations achieved in quantum metrology
where the concept of ${\cal PT}-$symmetry reemerged as
a balance between an engineered
gain and loss.
This led to multiple realizations of
unconventional phase transitions \cite{Rotter,Ri,Rii,Rii}.
In addition,
also the physics beyond the
critical points acquired certain new interpretations, say,
in the information-geometric framework  \cite{Riii,Rv}.
Last but not least,
the innovation of the theory encouraged
a wave of designs of multiple non-traditional physical devices
\cite{xi,xii,xiii,xiv,xv,xvi,xvii} (cf. also
further references therein).}

{After this detour to a broader
methodical as well as phenomenological context, we will,
nevertheless,
return, in what follows, to the mere
closed-system dynamical scenarios as sampled
by our example (\ref{eq:dimer-h})
possessing the spectrum of energies $E = \pm \omega$
which are both real.}

\section{Revisiting the Dyson's model-building strategy
through an inversion of the map}\label{ivni}

The simplicity of the example outlined above allows one to revisit
the
philosophy and,
in the spirit of the early review \cite{Geyer},
to invert the map (\ref{simn}).
With the Hamiltonian (\ref{eq:d}) now introduced {\it a priori},
we would like to emphasize the distinction, so we will denote
the preselected candidate for the Hamiltonian by a dedicated
tilded upper-case symbol $\widetilde{H}$.

Once we adopt this inverse-Dyson model-building strategy, the
first task is to verify that the eigenvalues
of the preselected $\widetilde{H}$
are real.
Fortunately, in our example, this is immediately seen to hold for
$|\gamma| < \kappa$. At the same time, the condition $|\gamma| <
\kappa$ ensures that $\alpha$ is real, so that both the
non-unitary Dyson-map matrix (\ref{eq:Omega-dimer}) and the
associated inner-product metric (\ref{HermiD}) are invertible.

One of the characteristic features of any``tilded-Hamiltonian''
generalization of the
theory is that the
construction of models
does not begin with Hamiltonian
$\mathfrak{h}$ or with the map (\ref{simn}). Instead, one
assumes that the (tilded) non-Hermitian (or, more precisely,
quasi-Hermitian) upper-case candidate $\widetilde{H}$ for the
Hamiltonian is given directly; see, for example, the
comprehensive reviews
\cite{Geyer} or \cite{Ali} for an explanation of
such a quasi-Hermiticity-based
approach.

The resulting updated theoretical framework may be viewed as an
eligible equivalent reformulation of the standard textbook quantum
mechanics. In this reformulation, typically, the
physics-motivated ansatz
(\ref{decoclass}) is replaced by its tilded analogue
 \be
 \widetilde{H}= \widetilde{H}_{\rm (free\ motion)}
 + \widetilde{H}_{\rm (non-Hermitian\ interaction)}
 \neq \widetilde{H}^\dagger\,,
 \label{undecoclass}
 \ee
i.e., by its non-equivalent,  mathematically motivated alternative.

An illustrative example of the change of paradigm can be found, for
instance, in Ref. \cite{BG}. Buslaev and Grecchi (BG)
studied there a truly remarkable non-Hermitian ordinary-differential
``minus-quartic'' anharmonic oscillator model, employing the
conventional kinetic-energy term
 \ben
 \widetilde{H^{\rm (BG)}}_{\rm (free\ motion)}
 =-\frac{1}{2}\frac{d^2}{dx^2}\,
 \ \
 \een
complemented by an asymptotically repulsive local
interaction,
 \be
 \ \ \
 \widetilde{H^{\rm (BG)}}_{\rm (non-Hermitian\ interaction)}(g,j,\epsilon)
 =\frac{j^2-1}{8(x-{\rm i}\epsilon)^2}
  +\frac{1}{2}(x-{\rm i}\epsilon)^2
 -g^2(x-{\rm i}\epsilon)^4\,.
 \label{BuGr}
 \ee
Regarding this truly anomalous toy model, which is constructed in
the entirely conventional Hilbert space $L^{2}(\mathbb{R})$, the
authors
were the first who
emphasized that their oscillator is time-reversal-parity
symmetric ($\mathcal{TP}$-symmetric; see Remark~4 in Ref.
\cite{BG}),
i.e., in the equivalent but more popular above-mentioned newer terminology,
$\mathcal{PT}$-symmetric.

In Ref. \cite{BG}, Buslaev and Grecchi also constructed, in a
closed and $\epsilon$-independent form, one of the self-adjoint
isospectral tilded twins $\widetilde{\mathfrak{h}^{\rm (BG)}}(g,j)$
of $\widetilde{H^{\rm (BG)}}_{}(g,j,\epsilon)$
which turned
out to be compatible with the
conventional textbook requirement~(\ref{decoclass}) (see the paper
for details). Their tilded-Hamiltonian model remains
exceptional, since, naturally, the simultaneous knowledge of the
initial Hamiltonian and of its Hermitian isospectral twin having the
conventional
form of Eq.~(\ref{decoclass}) resolves all questions regarding the
correct physical probabilistic interpretation of the underlying
quantum system.

Nontrivial and challenging interpretational questions arise only
in the study of systems for which the resulting Hermitian
$\widetilde{\mathfrak{h}}$ becomes complicated (i.e., typically, non-local).
In the
present paper, we therefore focus on such dynamical
scenarios and non-exceptional, generic models
in which one must work directly
with $\widetilde{H}$,
since its Hermitian twin $\widetilde{\mathfrak{h}}$ is far from being
user-friendly \cite{Ali, Batal}.
Accordingly, our present results
can be viewed as an explicit clarification of the problem of the
standard probabilistic interpretation of experiments within the
framework of the general tilded-Hamiltonian theory.

Several complementary remarks on tilded Hamiltonians and on the
associated quasi-Hermitian quantum mechanics (QHQM) as reviewed in
Refs. \cite{Geyer,Ali} are now in order. First, it should be noted
that the new form of the input information about
the unitary quantum dynamics implied that
the users
of the tilded-Hamiltonian models
had to provide a rigorous---and
often nontrivial---proof that the bound-state energy eigenvalue
spectra are real. For many operators
$\widetilde{H}$, such proofs required the application of
sophisticated mathematical techniques (cf., e.g., examples in Ref.
\cite{book}). Moreover, even the formally
real spectrum can happen to be fragile or,
at least, strongly
sensitive to the small changes of parameters \cite{fragile}.

Secondly, instead of the correspondence given in
Eq.~(\ref{simn}), one should rather refer to the Buslaev-inspired
\cite{VG} inverted reconstruction,
 \be
 \widetilde{H} \ \to \ \widetilde{\mathfrak{h}}= \Omega\,
 \widetilde{H}\,\Omega^{-1}
 =\widetilde{\mathfrak{h}}(\Omega)
 \,.
 \label{simnB}
 \ee
This brings us
back to the textbook framework, with the key advantage that such a
map restores the path toward the standard probabilistic
interpretation of all general observability aspects, both for
bound states (in the case of discrete spectrum) and/or for
scattering states (cf. Refs. \cite{Jones,discrete}).

Thirdly, the transition from the
physics-motivated
ansatz~(\ref{decoclass}) to its mathematics-motivated
analogue~(\ref{undecoclass})
has narrowed the
phenomenological
scope of the theory, especially
for a generic initial  $\widetilde{H}$
chosen, typically, in the mere ordinary differential-operator
form. In parallel, such a reduction of the class of models
encouraged
a perceivable intensification of the study of related
mathematics \cite{book,Violab,Viola}.

\section{Hiddenly Hermitian Hamiltonians: An
example \label{ni}}

For illustrative purposes, let us now recall the QHQM model of
Ref. \cite{Bijan}, in which the initial introduction of the
primary, manifestly non-Hermitian (i.e., tilded, upper-case)
Hamiltonian $\widetilde{H}$
was perceived as a fermionic analogue
of the widely studied bosonic Swanson oscillator
\cite{Swanson}. Its Hamiltonian
\begin{equation}
\label{eq:HF-fermionic}
\widetilde{H} = \omega_1 c_1^\dagger c_1 + \omega_2 c_2^\dagger c_2
    + \beta c_1^\dagger c_2^\dagger + \alpha c_2 c_1
\end{equation}
with positive $\omega_{1,2} > 0$ and
with the two real parameters $\alpha \neq
\beta$ (so that $H \neq H^\dagger$) involves the fermionic
annihilation operators $c_1$ and $c_2$ which satisfy the canonical
anti-commutation relations,
\begin{equation}
\{c_j,c_k^\dagger\} = \delta_{jk},
\qquad
\{c_j,c_k\} = 0 = \{c_j^\dagger,c_k^\dagger\},
\qquad j,k = 1,2.
\end{equation}
This model can be viewed as a two-site truncation of the Kitaev chain
\cite{Kitkit,Kitkitb} in the absence of inter-site
hopping but with superconducting-type pairing
interactions made non-Hermitian by the choice of distinct real
$\alpha\neq\beta$.

Denoting the vacuum state by $|0\rangle$, we have
\begin{equation}
c_1|0\rangle = 0,
\qquad
c_2|0\rangle = 0.
\end{equation}
The fermionic Fock space decomposes as
\begin{equation}
\mathcal{H}^{(4)}
 = \mathcal{H}_0^{(1)}\oplus\mathcal{H}_1^{(2)}\oplus\mathcal{H}_2^{(1)}
\end{equation}
with
\begin{equation}
\mathcal{H}_0^{(1)} = \mathrm{span}\{|0\rangle\}\,,\ \ \
\mathcal{H}_1^{(2)} = \mathrm{span}\{c_1^\dagger|0\rangle, c_2^\dagger|0\rangle\}\,,\ \ \
\mathcal{H}_2^{(1)} = \mathrm{span}\{c_1^\dagger c_2^\dagger|0\rangle\}\,.
\end{equation}
Introducing the basis
{$\{|1\rangle,|2\rangle,|3\rangle,|4\rangle
\}$
such that}
\begin{equation}
|1\rangle = |0\rangle,\quad
|2\rangle = c_1^\dagger|0\rangle,\quad
|3\rangle = c_2^\dagger|0\rangle,\quad
|4\rangle = c_1^\dagger c_2^\dagger|0\rangle,
\end{equation}
and choosing
\begin{equation}
\omega_1 = \omega,\qquad \omega_2 = 1-\omega,
\qquad \omega\in(0,1),
\end{equation}
a straightforward application of the anti-commutation relations yields
\begin{equation}
\widetilde{H}|1\rangle = \beta |4\rangle\,,\ \ \
\widetilde{H}|2\rangle = \omega |2\rangle\,,\ \ \
\widetilde{H}|3\rangle = (1-\omega) |3\rangle\,,\ \ \
\widetilde{H}|4\rangle = \alpha |1\rangle + |4\rangle.
\end{equation}
In this basis, the $4\times 4$ matrix representation of
$\widetilde{H}$ is
\begin{equation}
\label{eq:H-matrix-final}
\widetilde{H}=
\left (
\begin{array}{cccc}
 0 & 0 & 0 & \alpha  \\
 0 & \omega  & 0 & 0 \\
 0 & 0 & (1-\omega)  & 0 \\
 \beta  & 0 & 0 & 1 \\
\end{array}
\right )\,
\end{equation}
which is non-Hermitian whenever $\alpha\neq\beta$, even though all
parameters are real.

In view of Eq.~(\ref{simnB}), we now seek a Hermitian Hamiltonian
$\widetilde{\mathfrak{h}}$ and a Dyson map $\Omega$ such that
\begin{equation}
\label{eq:fermionic-Dyson-relation}
\widetilde{H} = \Omega^{-1} \widetilde{\mathfrak{h}} \Omega\,.
\end{equation}
Assuming $\alpha \beta > 0$, one can readily verify that one of
the valid isospectral twins of $\widetilde{H}$
is 
\begin{equation}
\label{eq:fermionic-h}
\widetilde{\mathfrak{h}} =
\left (
\begin{array}{cccc}
 0 & 0 & 0 & \sqrt{\alpha\beta} \\
 0 & \omega  & 0 & 0 \\
 0 & 0 & (1-\omega)  & 0 \\
 \sqrt{\alpha\beta} & 0 & 0 & 1 \\
\end{array}
\right ).
\end{equation}
This matrix is Hermitian and comprises a coupled two-level
subsystem in the $\{|1\rangle,|4\rangle\}$ sector with real
coupling $\sqrt{\alpha \beta}$, along with two decoupled
eigenstates, $|2\rangle$ and $|3\rangle$, having eigenvalues
$\omega$ and $1-\omega$, respectively.

Among the available
Dyson maps, one can now be particularly easily constructed
in its inverse-matrix form
\begin{equation}
\label{eq:Omega-inverse}
\Omega^{-1} =
\left (
\begin{array}{cccc}
 1 & 0 & 0 & (\alpha -\sqrt{\alpha\beta}) \\
 0 & 1 & 0 & 0 \\
 0 & 0 & 1 & 0 \\
 (\sqrt{\alpha\beta}-\beta) & 0 & 0 & 1 \\
\end{array}
\right ).
\end{equation}
The determinant of $\Omega^{-1}$ is
\begin{equation}
\mathcal{D}
= \det(\Omega^{-1})
= 2\alpha\beta - (\alpha+\beta)\sqrt{\alpha\beta} + 1,
\end{equation}
so that $\Omega$ exists and is invertible, provided that
$\mathcal{D} \neq 0$. Noting also that this matrix is real, so
that $\Omega^\dagger = \Omega^{\mathsf T}$, a straightforward
calculation yields
\begin{equation}
\Theta(\alpha,\beta)=
\left (
\begin{array}{cccc}
\displaystyle\frac{(\beta-\sqrt{\alpha\beta})^2 + 1}{\mathcal{D}^2}
  & 0 & 0 & \displaystyle\frac{\beta-\alpha}{\mathcal{D}^2} \\
0 & 1 & 0 & 0 \\
0 & 0 & 1 & 0 \\
\displaystyle\frac{\beta-\alpha}{\mathcal{D}^2} & 0 & 0 &
\displaystyle\frac{(\alpha-\sqrt{\alpha\beta})^2 + 1}{\mathcal{D}^2}
\end{array}
\right )\,,
\end{equation}
i.e., the physical inner-product metric is given in a
closed, compact, and sparse-matrix form.


\section{Towards a classification of the Dyson-map operators}\label{esivni}

Given any ``input-information'' operator $\widetilde{H}$ within
the QHQM framework, we will henceforth drop the tildes. We will
also refrain from distinguishing between the directions of the
isospectrality-mapping arrows in the Dyson-inspired
Eq.~(\ref{simn}) and its inverse (\ref{simnB}) (cf.
Table~\ref{Pwe2}). It should be kept in mind that
what is assumed to be known in advance
is solely the quasi-Hermitian operator $H$ representing the
observable energy Hamiltonian.

\begin{table}[h]
\caption{Two alternative choices of the Hamiltonian} \vspace{0.21cm}
 \label{Pwe2}
\centering
\begin{tabular}{ccccc}
 \hline \hline
    symbol&
 definition
 &
 self-adjoint  &interpretation
 & example
   \\
 \hline
 $\mathfrak{h}$ &
 Eq.~(\ref{decoclass})
 &
 yes
 &   conventional & Dyson \cite{Dyson}
 \\
 $H$&
 Eq.~(\ref{undecoclass})
 &
 no& via  twin $\mathfrak{h}$&Buslaev \cite{VG}
 \\
 \hline \hline
\end{tabular}
\end{table}

\subsection{The search for special, diagonal
 $\mathfrak{h}=\mathfrak{h}_I$}

If we restrict our attention to the
relationship between theory and measurement in closed quantum
systems that support $N$-plets of non-degenerate, stable bound
states (with $N$ finite or infinite, $N \leq \infty$), it becomes
sufficient to consider Schr\"{o}dinger equation
 \be
 H_{}\,|\psi^{}_n\kt=E_n\,|\psi^{}_n\kt\,,
 \ \ \ \ n = 1, 2, \ldots, N\,
 \label{unoSE}
 \ee
and, due to the non-Hermiticity $H \neq H^\dagger$, also its
conjugate version
 \be
 H^\dagger_{}\,|\psi^{}_n\kkt
 =E_n\,|\psi^{}_n\kkt\,,
 \ \ \ \ n = 1, 2, \ldots, N\,
 \label{duoSE}
 \ee
in which we used the notation convention of Ref.~\cite{SIGMA},
i.e., the double-ket symbol for the right eigenvectors of
$H^\dagger$ (and, if needed, also its double-bra alternative $\bbr
\psi^{}_n|\,$ for their conjugates).

In both of these equations, the normalization of the eigenvectors
may be chosen arbitrarily but kept fixed. Once fixed, the set of
all column vectors $|\psi_n\kkt$ can be concatenated and
interpreted as an $N \times N$ matrix, 
 \be
 \left \{
 |\psi^{}_1\kkt,|\psi^{}_2\kkt,
 \ldots,|\psi^{}_N\kkt
 \right \} := \Omega_I^\dagger\,.
 \label{konkate}
 \ee
Subsequently, a comparison of Eq.~(\ref{duoSE}) with Eq.
(\ref{simn})---or with Eq.~(\ref{simnB}), keeping in mind
that we now ignore
the tildes---reveals a remarkable conclusion: the matrix
$\Omega_I^\dagger$ of Eq.~(\ref{konkate}) is precisely the
Hermitian conjugate of the Dyson map that renders the
Hermitian avatar of $H$ diagonal.

The latter map
as well as the Hermitian avatar of $H$
may be
$I$-subscripted, so we have
 \be
 \left (\mathfrak{h}_I
 \right )_{nn}=E_n\,,
 \ \ \ \ n = 1, 2, \ldots, N\,,\ \ \ N \leq \infty\,.
  \ee
In this matrix, all of the eigenvalues of $H$, or equivalently of
$H^\dagger$, appearing on its diagonal are real.

\subsection{A broader class of Dyson maps}

The above construction makes it clear that the
resulting operator $\Omega_I$ is not unique. Many other
Dyson maps may exist. The reason is that, at every
index $n$, Eq.~(\ref{duoSE}) remains unchanged under
multiplication by any real or complex constant $\kappa_n \neq 0$.
So, a different concatenation (\ref{konkate}) would result.
Consequently, we may introduce a diagonal and invertible matrix
$K$ (whose diagonal entries are precisely these constants)
and define the product
 \be
 \Omega_K=K^\dagger\,\Omega_I\,,
 \label{[14]}
 \ee
i.e., the equally admissible Dyson-map operator, which is
parametrized by the
$K$ matrix (and, correspondingly, carrying the $K$-matrix
subscript). Thus, the
Dyson map $\Omega_I$
of the preceding paragraph
(carrying the $I$ subscript corresponding
to the identity matrix) can be viewed as the mere special $K = I$
case
of a more general $N$-parametric family~(\ref{[14]}).

Two immediate conclusions can be drawn. First, a
change in $K$ implies a change in the inner product
metric in general [cf. Eq.~(\ref{genn})],
 \be
 \Theta=
 \Theta_K=\Omega_K^\dagger\Omega_K=\Omega_I^\dagger
 K K^\dagger \Omega_I
 \neq \Theta_I =\Omega_I^\dagger\Omega_I\,.
 \label{[15]}
 \ee
Second, as long as
 \be
 \mathfrak{h}_K
 =K^\dagger \,\mathfrak{h}_I \left (K^\dagger \right )^{-1}
 =\mathfrak{h}_I\,,
 \label{[16]}
 \ee
none of the different $K$-subscripted metrics leads to a violation
of diagonality of the Hermitian-matrix twin
$\mathfrak{h}_I$ of our preselected non-Hermitian Hamiltonian
$H$.

\subsection{The most general form of the Dyson map}

In the light of the illustrative example of Buslaev and Grecchi
\cite{BG}, given by Eq. (\ref{BuGr}), it becomes clear that even
the most elementary diagonal matrix $\mathfrak{h}_K =
\mathfrak{h}_I$ need not, in fact, be the preferred choice for the
Hermitian isospectral twin  of $H$. Indeed, if any (not
necessarily diagonal) matrix $\mathfrak{h}$ is to serve as the
background for a meaningful probabilistic interpretation of the
quantum system in question, its optimal form may, in general, be
merely a suitable unitary transformation of $\mathfrak{h}_I$,
 \be
  \mathfrak{h}  = {\cal U}\,\mathfrak{h}_I\, {\cal U}^{\dagger}
  :=\mathfrak{h}{[\cal U]}\,,
  \ \ \ \ \
 {\cal U}^{-1}={\cal U}^\dagger\,.
   \label{unitas}
 \ee
The authors of Ref.~\cite{BG} provided a clear and explicit example
 \be
 \mathfrak{h}^{\rm (BG)}
 =-\frac{d^2}{dy^2} + y^2\,(gy-1)^2 -j\,(gy-1/2)\,
 \label{aha}
 \ee
of such an optimal, albeit non-diagonal, operator
$\mathfrak{h}{[\cal U]}$ which depends on a nontrivial ${\cal
U}\neq I$.

For our illustrative example in Eq.~(\ref{aha}), both the
probabilistic interpretation and the stable bound-state nature of
the eigenstates of this self-adjoint double-well Hamiltonian are
standard. Consequently, for conceptual and interpretational
purposes alone, diagonalizing operator (\ref{aha}) would be,
from the correct interpretation point of view, entirely
superfluous.

In similar cases, it may easily happen that the diagonalization
${\cal U} : \mathfrak{h} \to \mathfrak{h}_I$ is only approximate.
Again, for a physically meaningful interpretation of the system,
reconstructing ${\cal U}$ (which might be purely numerical) is
entirely redundant. Thus,
we may conclude that in addition to the constructions of the
Dyson map $\Omega = \Omega_I$ [cf. Eqs.~(\ref{konkate}) and
(\ref{duoSE})] and of its
$K-$reparametrized descendants $\Omega = \Omega_K$ [cf.
Eq.~(\ref{[14]})], it remains for us to consider
the third class of the Dyson maps
 \be
 \Omega_{K,\cal U}={\cal U}\,K^\dagger\,\Omega_I\,
 \label{[u14]}
 \ee
characterizing the physics behind models
with
nontrivial ${\cal U} \neq I$, i.e., with
both the $K-$ and ${\cal U}-$parametrized family
of the fully general Dyson maps (cf.
the ultimate classification scheme in Table~\ref{Qwe2}).

\begin{table}[h]
\caption{Three alternative
choices of variable parameters.} \vspace{0.21cm}
 \label{Qwe2}
\centering
\begin{tabular}{cccc}
 \hline
 \hline
%
parameters
 &
 Dyson map $\Omega$&
 Hermitization of $H$&metric $\Theta$
 \\
 \hline 
 -&$\Omega_I$
  &
 $
 \mathfrak{h}_I$&
 $\Theta_I$
 \\
 $K$&
 $K^\dagger\,\Omega_I=\Omega_K$
 &
 $
 \mathfrak{h}_I$
 &
 $\Theta_K$
 \\
 $K$, ${\cal U}$
 &
 ${\cal U}\,K^\dagger\,\Omega_I=\Omega_{K,\cal U}$
 &
 ${\cal U}\,\mathfrak{h}_I\, {\cal U}^{\dagger}=\mathfrak{h}[{\cal U}]$
 &
 $\Theta_K$
 \\
 \hline \hline
\end{tabular}
\end{table}


\section{Discussion\label{essiq}}

\subsection{\label{sigur}Complete sets of independent quasi-Hermitian observables}

A generic quantum model is usually characterized not only
by its Hamiltonian but also by a number of
additional
observables. For example, in Ref. \cite{Batal}, the authors
considered a
QHQM model
with a non-Hermitian
Hamiltonian $H \neq H^\dagger$
and
with a non-Hermitian
particle-position operator $X \neq X^\dagger$. Although the necessity
and feasibility of similar extensions
of the theory
were already discussed by
Scholtz et al.~\cite{Geyer}, only a few implementations
of such an idea appear in
the literature.

The reasons
are far from obvious,
and not too many authors
have noticed that under the quasi-Hermiticity constraint
(\ref{Dys}), the simultaneous $H-$ and $K-$dependence of the
inner-product metric $\Theta$ [cf. Eqs.~(\ref{duoSE}),
(\ref{konkate}), and (\ref{[15]})] strongly limits its
compatibility with any other, independently chosen (preselected)
non-Hermitian operator $A \neq A^\dagger$ having real spectrum
\cite{arabky}.
Even the guarantee of compatibility of
two preselected non-Hermitian observables $H$ and $A$ (with
real spectra)
is a nontrivial task
requiring, in general, the full freedom and generality
of the $K-$ and ${\cal U}-$ dependent
Dyson map as represented by Eq.~(\ref{[u14]})
(cf. also
the last line of
Table~\ref{Qwe2}).

By construction, the unitary-matrix factor ${\cal U}$ mediates
only the diagonalization of the Hamiltonian $\mathfrak{h}$, but
generally not that of the $A$-associated twin $\mathfrak{a}$.
Consequently, for a fixed ${\cal U}$, any other linearly
independent and admissible non-Hermitian (more precisely,
$\Theta_K$-quasi-Hermitian) candidate $A$ for an
independent observable must belong to a rather restricted
family of eligible (i.e., probabilistically interpretable)
Hermitian avatars
 \be
 {\mathfrak{a}}= {\cal U}\,K^\dagger\,\Omega_I\,
 {A}\,\Omega_I^{-1}\,
 \left ( K^\dagger
 \right )^{-1}\,{\cal U}^{-1}\,.
 \label{simnA}
 \ee
They are, in general, manifestly $K-$ and
${\cal U}-$dependent,
 $$
  {\mathfrak{a}}= {\mathfrak{a}}_K[{\cal U}]\,.
 $$
All of this holds unless the central matrix factor $\Omega_I A
\Omega_I^{-1}$ in (\ref{simnA}) is diagonal; that is,
unless all of the eigenstates of
$A^\dagger$ coincide with the eigenstates $|\psi_n\kkt$ of
$H^\dagger$ as determined by the conjugate Schr\"odinger equation
(\ref{duoSE}).

One can conclude that the
general Hermitian Hamiltonian (\ref{unitas}) remains independent
of $K$ [cf. Eq.~(\ref{[16]})], while the general metric operator,
due to Eq.~(\ref{[u14]}), remains independent of the unitary
matrix ${\cal U}$. Nevertheless, the two-matrix dependence
(\ref{[u14]}) of the general Dyson map remains crucial, as it
enlarges the class of admissible pairs of
independent observables $H$ and $A$.

%
%
%

\subsection{Models in which the Dyson map is required Hermitian\label{espiq}}

The guarantee of compatibility between $H$ and an
independent $A$ becomes trivial if we first choose $H$, construct
a suitable metric $\Theta = \Theta(H)$, and then define the class
of admissible $A$ operators via the relation
 \be
 A^\dagger\,\Theta = \Theta\,A={\cal M}={\cal M}^\dagger
 \ee
in which the $N \times N$ Hermitian matrix ${\cal M}$ (with $N
\leq \infty$) is treated as an input choice.
This allows one to
define
all of the eligible
observable $A=A({\cal M})$ by the following explicit
matrix-multiplication formula
 \be
 A({\cal M})=\Theta^{-1}\,{\cal M}\,
 \ee
in which the Hermitain matrix of parameters ${\cal M}$ is arbitrary.

In many applications,
the theory is often complemented by
an additional
(and, in fact, entirely formal)
requirement of having
the
Dyson map defined in the specific
form of a Hermitian square root
of the metric, i.e.,
such that $\Omega = \Omega^\dagger$:
for a sample of such a
purely mathematically motivated auxiliary postulate see
Eq.~(\ref{HermiD}) above.
Within the context of our exhaustive classification of
$H$-compatible Dyson maps $\Omega = \Omega(H)$ (cf.
Table~\ref{Qwe2}), such an
optional
Hermiticity condition
imposes, in effect, a fairly nontrivial
constraint upon the parameters,
i.e., upon all of the matrix elements of ${\cal U}$
and, as a consequence,
upon the class of the other, complementary eligible
observables $A=A({\cal M})$.
At the same time,
the Hermiticity condition $\Omega =
\Omega^\dagger$, {i.e.,
 $$
 {\cal U}\,K^\dagger\,\Omega_I=\Omega_I^\dagger\,K\,{\cal U}^\dagger
 $$}
remains tractable and
can be
satisfied when
re-expressed in its $N \times N-$matrix
quadratic-equation form
 \be
 {\cal U}\,\Omega_K\,{\cal U}
 =\Omega^\dagger_K
 \,.
 \label{[w14]}
 \ee
The number of the unknown (real) parameters equals the number of the
independent scalar constraints,
so that the equation can generally be solved
and
specify the unitary matrix ${\cal U}$, in principle at least.
Even in practice the recipe still seems to lead to a feasible
construction, especially when
the Hilbert-space dimension $N$ remains finite and is not too large.
Again, an elementary example of the result can be found here
in section \ref{vni} above.


\section{Summary\label{progresivni}}

The Dyson-inspired non-Hermiticity-admitting
simplification
$\mathfrak{h} \to H \neq H^\dagger$
of a  preselected Hermitian Hamiltonian
[cf.
Eq.~(\ref{simn})] is, from
a purely physical and phenomenological point of view, trivial.
In contrast, a truly nontrivial physical interpretation problem
arises when one starts from a non-Hermitian $H$
with a real spectrum.
Indeed, the entirely formal decomposition
of
Eq.~(\ref{simnB})
can hardly be given a sound observable meaning
without a clear
reconstruction of its isospectral Hermitian twin.

In the present paper, we formulated and addressed this problem. We
proposed that, in general, the reconstruction of $\mathfrak{h}$
can be classified as summarized in Table~\ref{Qwe2}. In this
scheme, increasing the flexibility of the process---that is,
increasing the number of available free parameters that determine
the complexity of the Dyson map and of the Hermitian avatar
$\mathfrak{h}$ of $H$---naturally leads to a corresponding
increase in the complexity of the correct physical
probabilistic interpretations of the quantum system in question.

Our systematic analysis of the underlying mathematical structures
was complemented by several illustrative examples, in which we
emphasized the balance between the complexity of the model and its
constructive tractability. On this basis, we conclude that once
the parameters (i.e., the two finite or infinite matrices $K$ and
${\cal U}$) are explicitly controlled, as ensured by our triple
Dyson-map classification, the QHQM formalism is well suited to
accommodate a wide range of current and future realistic
applications.

\newpage

\appendix
\renewcommand{\theequation}{A\arabic{equation}}
\setcounter{equation}{0} 


\section{Introduction to Dyson-map theory `for pedestrians':
Definition, fundamental properties, and terminology}

For the sake of clarity---especially for non-expert readers---we
briefly recall the core definition and essential properties of the
operators and terminology used throughout the present
paper.
Such an addendum provides an alternative summary of the QHQM theory
presented from a slightly different, complementary
point of view.
For the sake of definiteness, the non-expert readers
might assume that
the Hilbert-space dimension is finite, $N<\infty$.
Still,
the word of warning is that only
some of the related observations
can readily be extended to hold at $N=\infty$ as well.
In the latter case, indeed, the use of a
more subtle mathematics may prove unavoidable
(see, e.g., \cite{book}).

In a finite$-N$
reformulation of the theory, the \emph{Dyson map} is \emph{any} linear
and
invertible operator that
``Hermitizes''
a non-Hermitian Hamiltonian. Concretely,
under certain purely technical assumptions concerning the
given non-Hermitian Hamiltonian $H$ with a real spectrum,
a Dyson
map $\Omega$ satisfies Eq.~(\ref{simn}) as well as its inversion
\begin{equation}
  \mathfrak{h} \;=\; \Omega\,H\,\Omega^{-1}
    \label{Eq_A1}
\end{equation}
with this operator
required to be Hermitian, $\mathfrak{h} \;=\;
\mathfrak{h}^\dagger$. The Dyson map defines the physical inner
product (metric) in the original Hilbert space ${\cal H}^{(N)}$ by
formula (\ref{genn}), i.e.,
\begin{equation}
  \Theta \;=\; \Omega^\dagger \Omega\,.
  \label{Eq_A2}
\end{equation}
This corresponds to the fact that $H$ is
$\Theta$-quasi-Hermitian (cf. Eq.~(\ref{Dys})):
\begin{equation}
  H^\dagger \, \Theta \;=\; \Theta \, H .
  \label{Eq_A3}
\end{equation}
Thus, the essential properties of the QHQM theory may be
summarized as follows.

\begin{itemize}

\item
 Hermitization: $\Omega$ establishes an
isospectral equivalence between the non-Hermitian $H$ and the
Hermitian twin $\mathfrak{h}$, hence both share the same energy
spectrum.

\item
 Unitary evolution: Time evolution generated by $H$
is unitary with respect to the inner product defined by $\Theta$.

\item
Observables: An operator $A$ is a physical observable in the same
representation iff it is $\Theta$-quasi-Hermitian: $A^\dagger
\Theta = \Theta A$. Equivalently, its Hermitian avatar is
$\mathfrak{a}=\Omega A \Omega^{-1}$.

\item
Non-uniqueness: The Dyson map is not unique; different choices
of $\Omega$ may yield different inner-product metrics
$\Theta=\Theta_K$ and therefore
different physical interpretations
of the  unitary quantum system in question.

\end{itemize}

As we have emphasized in the main text, different constructions of
$\Omega$  lead to non-equivalent
probabilistic interpretations of the observables $H$ and $A$, etc.
More explicitly, this
is an observation which
has been discussed in section \ref{essiq} and
summarized in Table \ref{Qwe2}.
It also underlies the classification of admissible
Dyson maps and metrics. Naturally, this is a formal ambiguity
with many immediate consequences in applications. {\it Pars pro toto},
let us mention that
recently,
the concept of the Dyson map was also
proposed to play the role of a (generalized)
vielbein  \cite{Ju2022}, highlighting the connection between the
QHQM
formalism and the standard vielbein approach widely used in many
areas of physics, including general relativity, supergravity, and
superstring theory.

Remark on terminology: In this paper we use the terms Hermitian
avatar and isospectral twin of a non-Hermitian Hamiltonian $H$.
Specifically, a \emph{Hermitian avatar} of $H$ is the Hermitian
operator $\mathfrak{h}$ defined by Eq. (A1
) with a Dyson map
$\Omega$. It represents the same quantum system as $H$, but
expressed in the standard Hilbert-space inner product. Moreover,
an \emph{isospectral twin} emphasizes that $H$ and $\mathfrak{h}$
share the same spectrum, $\mathrm{spec}(H) =
\mathrm{spec}(\mathfrak{h})$, and are related by a (typically,
non-unitary) similarity transformation. Thus the Hermitian avatar
$\mathfrak{h}$ is simultaneously the \emph{isospectral twin} of
the non-Hermitian Hamiltonian $H$.

\subsection*{Funding:}

A.M. was supported by the Polish National Science Centre (NCN)
under the Maestro Grant No. DEC-2019/34/A/ST2/00081.

\subsection*{Data availability statement:}

No new data were created or analysed in this study.

\subsection*{Conflicts of Interest:}

The authors declare no conflicts of interests.

\subsection*{ORCID ID:}

  \noindent
Aritra Ghosh: https://orcid.org/0000-0003-0338-2801

  \noindent
Adam Miranowicz: https://orcid.org/0000-0002-8222-9268

  \noindent
Miloslav Znojil: https://orcid.org/0000-0001-6076-0093

\end{document}